\documentclass[12pt,preprint]{aastex}

\begin{document}
\title{On the Insignificance of Photochemical Hydrocarbon Aerosols in the Atmospheres of Close-in Extrasolar Giant Planets}

\author{Mao-Chang Liang$^{1}$, Sara Seager$^{2}$, Christopher D. Parkinson$^{1,3}$, Anthony Y.-T.  Lee$^{1}$, and Yuk L. Yung$^{1,3}$}
\affil{$^{1}$Division of Geological and Planetary Sciences, California Institute of Technology, 1201 E. California Blvd., Pasadena, CA 91125}
\affil{$^{2}$Department of Terrestrial Magnetism, Carnegie Institution of Washington, 5241 Broad Branch Rd. NW Washington, D.C. 20015}
\affil{$^{3}$Jet Propulsion Laboratory/California Institute of Technology and the NASA Astrobiology Institute}

\begin{abstract}
The close-in extrasolar giant planets (CEGPs) reside in irradiated
environments much more intense than that of the giant planets in
our solar system. The high UV irradiance strongly influences their
photochemistry and the general current view believed that this
high UV flux will greatly enhance photochemical production of
hydrocarbon aerosols. In this letter, we investigate hydrocarbon
aerosol formation in the atmospheres of CEGPs. We find that the
abundances of hydrocarbons in the atmospheres of CEGPs are
significantly less than that of Jupiter except for models in which
the CH$_4$ abundance is unreasonably high (as high as CO) for the
hot (effective temperatures $\gtrsim 1000$~K) atmospheres.
Moreover, the hydrocarbons will be condensed out to form aerosols
only when the temperature-pressure profiles of the species
intersect with the saturation profiles---a case almost certainly
not realized in the hot CEGPs atmospheres. Hence our models show
that photochemical hydrocarbon aerosols are insignificant in the
atmospheres of CEGPs. In contrast, Jupiter and Saturn have a much
higher abundance of hydrocarbon aerosols in their atmospheres
which are responsible for strong absorption shortward of 600~nm.
Thus the insignificance of photochemical hydrocarbon aerosols in
the atmospheres of CEGPs rules out one class of models with low
albedos and featureless spectra shortward of 600~nm.
\end{abstract}

\keywords{planetary systems---radiative transfer---stars:
atmosphere---stars: individual (HD~209458)}

\section{Introduction}

Hazes and clouds\footnote{``Hazes" refers to the diffuse and
optically thin aerosol distribution, while ``clouds" refers to the
optically thick regions \citep{Wetal86}.} in the atmospheres of
jovian planets can strongly affect the ability to determine
atmospheric composition at ultraviolet to infrared wavelengths. At
wavelengths shorter than $\sim$600~nm, the atmospheric line
features in the jovian planets are ``washed out" by the
hazes/clouds in the atmospheres of planets (e.g., Karkoschka \&
Tomasko 1993; Karkoschka 1998). The main chemical compositions of
the hazes/clouds on Jupiter are believed to be H$_2$O-NH$_3$,
NH$_4$SH, NH$_3$, N$_2$H$_4$, and hydrocarbons from several bar to
$\sim$0.1~mbar
\citep{WL73,S83,Wetal86,PH91,Getal96,Betal98a,Betal98b,Wetal03}.
Saturn may have a composition profile similar to Jupiter since
they have similar 300-1000~nm spectra (e.g., Karkoschka 1998).
Saturn's albedo has been successfully modelled by assuming a
dichotomy in the aerosol distribution between the troposphere and
stratosphere, where the number density of aerosols is much lower
in the stratosphere \citep{KT93}. It is found that the
stratospheric aerosols are very dark at $\sim$300~nm, implying the
presence of hydrocarbon aerosols.

Since the recent increase in sample size of extrasolar planets
(e.g., Udry et al. 2002; Butler et al. 2003), the planetary
formation environment has been statistically analyzed, although
not conclusively \citep{Fetal02,SIMRU03}. The close-in extrasolar
giant planets (CEGPs, with semi-major axes $\lesssim 0.05$ AU,
also known as ``hot Jupiters") are of particular interest since
they have more active chemical processes in their atmospheres
(e.g., Liang et al. 2003) and the evolution of the atmospheres can
currently be studied observationally (e.g., Vidal-Madjar et al.
2003, 2004). A number of simulations in the atmospheres of CEGPs
have been performed to study the albedos and reflection spectra by
including the formation of high temperature condensates, such as
silicates (e.g., Sudarsky et al. 2000; Seager et al. 2000).  The
importance and existence of the atmospheric aerosols have been
addressed and discussed widely in recent years (e.g., Baraffe et
al. 2003) and it is generally believed that more UV flux will
result in more aerosols. The photochemistry in jovian atmospheres
results in photochemical aerosols which significantly affect the
ultraviolet-visible spectra and albedos; hence we were motivated
to simulate the formation of various molecules, e.g.,
hydrocarbons, ammonia, and sulfuric acid, which are the possible
sources of aerosols, in the atmospheres of CEGPs. In this letter,
we focus on hydrocarbons and hydrocarbon aerosol formation.

\section{Model}
\begin{figure*}
\epsscale{1}
\plotone{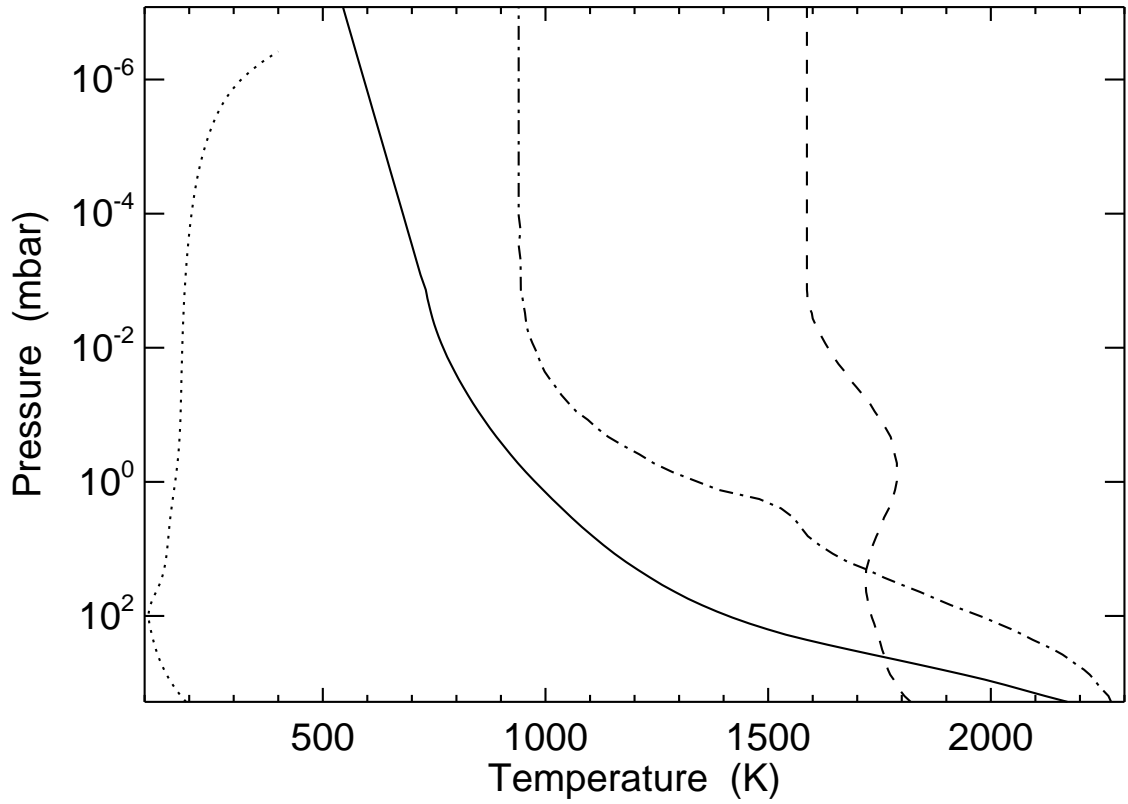} \caption[3 modelled T-P
profiles]{Vertical temperature profiles of the reference model
(solid line), \citet[dashed line]{Betal02}, \citet[dash-dotted
line]{Fetal03}, and Jupiter (dotted line).
We assume the profiles of \citet{Betal02} and \citet{Fetal03} are
isothermal above their reported pressure levels. \label{profile}}
\end{figure*}

A one-dimensional Caltech/Jet Propulsion Laboratory KINETICS model
is applied to HD~209458b's atmosphere, which is divided into 80
plane-parallel layers along the radial direction. The planet is
probably tidally locked and our simulation is performed on the day
side. The model assumes the four parent molecules$:$ H$_2$, CO,
H$_2$O, and CH$_4$. The abundances of CO and H$_2$O for the
reference model (Model A) are $3.6 \times 10^{-4}$ and $4.5 \times
10^{-4}$, respectively. The CH$_4$ abundance is taken to be $3.9
\times 10^{-8}$, which is the low bound predicted by \citet{SS00}.
The temperature-pressure profiles are not certain, because the
global circulation and high temperature condensation are not
constrained in generating the model atmosphere. Our reference
profile (solid curve in Figure~\ref{profile}) is a derivative of a
cloud-free and high temperature condensation-free model. The
stellar irradiance is assumed to be uniformly distributed over the
whole planet; this gives the lower bound of the temperature
profile in the atmosphere of HD~209458b. In view of the
aforementioned uncertainty, two alternative temperature profiles,
which assume the redistribution of the stellar irradiance evenly
only on the day side, are examined \citep{Betal02,Fetal03}.

A one-dimensional, photochemical-diffusive, diurnally averaged
numerical model for hydrocarbon photochemistry has been presented
by \citet{Getal96} in the atmosphere of Jupiter.  In that study,
important chemical cycles and pathways involving C$_1$-C$_{4}$
species are identified. Included in this analysis are sensitivity
studies on a standard reference model with respect to variations
in the eddy-diffusion profile, solar flux, atomic hydrogen influx,
latitude, temperature, and important chemical reaction rates. The
model reproduces extensive observations of hydrocarbon species as
well as He 584 \AA\ and H Lyman-$\alpha$ airglow emissions on
Jupiter. Due to the incomplete laboratory measurements of reaction
rates and photodissociation quantum yields in the C$_3$ and higher
hydrocarbons, we use a simplified version of the hydrocarbon
photochemical model by \citet{Getal96}. The hydrocarbon chemistry
up to the C$_2$ hydrocarbons is modelled thoroughly in the
atmosphere of HD~209458b. The C$_1$ and C$_2$ hydrocarbons are the
fundamental ingredient for building up complex hydrocarbons, e.g.,
benzene and polycyclic aromatic hydrocarbons (PAHs), through long
chain polyynes and polymerization. The chemical pathways among the
C$_1$ and C$_2$ in the atmospheres of CEGPs were first pointed out
by \citet{Letal03}, which are fundamentally different from the
pathways on the colder jovian planets \citep{Getal96}. The full
version of the hydrocarbon photochemical model is also verified.
The oxygen related photochemistry is taken from
\citet{Mosesetal00}.

Figure~\ref{profile} shows the temperature profiles for three
models \citep{SS00,Betal02,Fetal03}. For each case, we have
examined five different initial chemical abundances for CH$_4$,
CO, and H$_2$O as tabulated in Table~\ref{abundance}. Due to the
unconstrained CH$_4$ abundance, we have varied it by several
orders of magnitudes to study its sensitivity in the formation of
hydrocarbons. However, we expect CO to be the dominant reservoir
of carbon for the range of temperatures in the atmospheres of
CEGPs and assume this in our reference Model~A. The models of
\citet{Betal02} and \citet{Fetal03} go only to 1 and 0.1~$\mu$bar
pressure levels, respectively$:$ we assume the profiles are
isothermal above these pressure levels. The parameters for the
reference eddy-diffusion profile ($\kappa = \kappa_0
(n/n_0)^{-\alpha}$, where $n$ is number density) are taken to be
$\kappa_0 = 2.4\times 10^{7}$~cm$^2$~s$^{-1}$, $n_0 = 5.8\times
10^{18}$~cm$^{-3}$, and $\alpha = 5.6$. We also varied $\kappa_0$
and $\alpha$ to test the sensitivity of the results on
eddy-diffusion (see Table~\ref{column}). The fiducial
eddy-diffusion used here is consistent with the upper limit
estimates from \citet{SG02}.

\clearpage

\clearpage

\begin{deluxetable}{rrrr}
\tabletypesize{\scriptsize} \tablecolumns{4} \tablecaption{Initial
Chemical Abundances of CH$_4$, CO, and H$_2$O for Models A-E.
\label{abundance}} \tablewidth{0pt} \tablehead{
\multicolumn{1}{c}{Model} & \multicolumn{1}{c}{CH$_4$} &
\multicolumn{1}{c}{CO} & \multicolumn{1}{c}{H$_2$O} }
\startdata
A        &$ 3.9\times 10^{-8}$ & $ 3.6\times 10^{-4}$ & $ 3.6\times 10^{-4}$ \\
B        &$ 3.9\times 10^{-8}$ & $ 3.6\times 10^{-4}$ & $ 3.6\times 10^{-5}$ \\
C        &$ 3.9\times 10^{-8}$ & $ 3.6\times 10^{-5}$ & $ 3.6\times 10^{-4}$ \\
D        &$ 3.9\times 10^{-5}$ & $ 3.6\times 10^{-4}$ & $ 3.6\times 10^{-4}$ \\
E        &$ 3.6\times 10^{-4}$ & $ 3.6\times 10^{-4}$ & $ 3.6\times 10^{-4}$
\enddata
\end{deluxetable}

\clearpage

\section{Results}

\begin{figure*}
\epsscale{0.8} \plotone{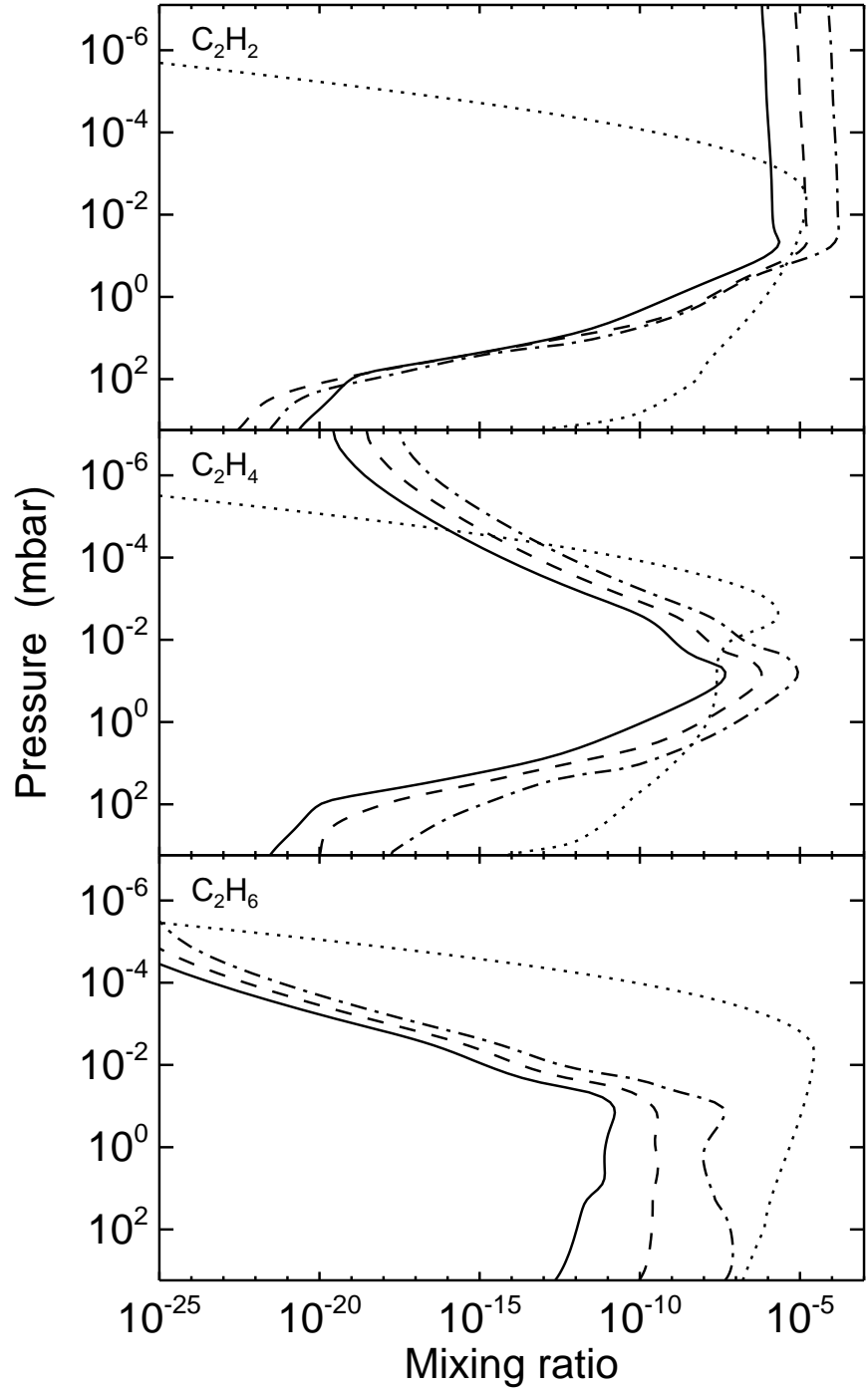} \caption[hydrocarbon mixing
ratios]{Comparison of volume mixing ratios of C$_2$H$_2$ (upper
panel), C$_2$H$_4$ (middle panel), and C$_2$H$_6$ (lower panel)
for Models A, D, E, and Jupiter (solid, dashed, dash-dotted, and
dotted lines, respectively). The high C$_2$H$_2$ mixing ratio at
the top of the atmosphere is due to the high photolysis rate of
CO. \label{mr}}
\end{figure*}

Our modeling shows that gas phase hydrocarbons are most likely
present in very low abundances in the atmospheres of CEGPs. This
result is in contrast to the high abundances of hydrocarbons on
the solar jovian planets. The vertical profiles of the
hydrocarbons for various models are shown in Figure~\ref{mr} and
the maximum and column integrated hydrocarbons are tabulated in
Table~\ref{column}. The hydrocarbons are produced and concentrated
mainly in the middle atmosphere, around 0.1~mbar. Because the
framework of hydrocarbon formation on the jovian planets is well
understood, we explain our results in comparison to the
photochemical production of hydrocarbons on the jovian planets.

There are two known chemical schemes for the formation of
hydrocarbons in the jovian atmospheres and their satellites. The
first is via the synthesis of long chain polyynes from C$_2$H$_2$
\citep{Aetal80}. The second is the polymerization of C$_2$H$_2$ to
form ring compounds \citep{Wetal00}. In both cases, C$_2$H$_2$
plays a crucial role. Therefore, to explain why hydrocarbon
aerosols are not formed in CEGPs, we have to explain why
C$_2$H$_2$ concentrations are so low. This is due primarily to the
high temperatures in the atmospheres of CEGPs and secondarily to
the high UV flux. Both the high temperatures and high UV fluxes
are a direct consequence of the CEGPs' closer proximity to their
parent stars.

One reason for low hydrocarbon abundances in CEGPs is because the
abundance of CH$_4$ is many orders of magnitudes lower than that
in the jovian atmospheres. The CH$_4$ abundance is important
because in the jovian atmospheres hydrocarbon formation is driven
by the photodissociation of CH$_4$ and the subsequent reactions of
the products (e.g., Gladstone et al. 1996).  The three species,
C$_2$H$_2$, C$_2$H$_4$, and C$_2$H$_6$, are important for forming
more complex hydrocarbons and hydrocarbon aerosols. The primary
reservoir of C in CEGPs is CO, not CH$_4$, as in the jovian
planets.  This is due to the much higher temperatures in the
atmospheres of CEGPs (effective temperatures $\gtrsim 1000$~K)
compared to Jupiter (effective temperature $\sim$130~K).
\citet{Letal03} showed that C compounds are initiated by C atoms
produced by the photolysis of CO in the upper atmosphere. The
hydrocarbons (C$_2$H$_2$, C$_2$H$_4$, and C$_2$H$_6$) are formed
along with CH$_4$ from the C atoms.

A second reason for the low abundance of hydrocarbons is that
hydrogenation of C$_2$H$_2$ to CH$_4$ by the pathways given in
\citet{Letal03} (see also Chapter~5 of Yung and DeMore 1999)
rapidly removes C$_2$H$_2$. As pointed out by \citet{Letal03}, the
CEGPs have a high concentration of H atoms formed via an H$_2$O
mediated process. Hydrogenation is the dominant removal process of
C$_2$H$_2$ in CEGPs and is driven by the high concentration of H
atoms. Unlike the colder jovian atmospheres, the hydrocarbon loss
via photolysis is minor in the atmospheres of CEGPs. A key
reaction in hydrogenation of C$_2$H$_2$ to CH$_4$ is the reaction
C$_2$H$_3$ + H$_2$ $\rightarrow$ C$_2$H$_4$ + H. The reaction that
breaks the H$_2$ bond is fast for the high temperatures in the
atmospheres of CEGPs; however in the colder atmospheres of the
jovian planets this reaction is the major bottleneck to
hydrogenation of C$_2$H$_2$. Hydrogenation as a cause of low
hydrocarbon abundances is therefore related to the high
temperatures in the atmospheres of CEGPs which are hot enough not
only for the rapid hydrogenation rate but also for H$_2$O to be
present in vapor form. In contrast to the jovian planets and their
satellites, water is frozen into ice and not available for
photolysis.

To show the robustness of the result of low hydrocarbon abundances
in the atmospheres of CEGPs, we varied the input parameters to our
photochemical model.  We find that over a broad range of input
parameters, i.e., initial chemical abundances and temperature and
eddy-diffusion profiles, the hydrocarbon formation in the
atmospheres of CEGPs never exceeds that of Jupiter. In our model
of an extremely abundant CH$_4$ (Model~E), the column integrated
hydrocarbon is about 0.5 that of Jupiter's (see
Table~\ref{column}). However, this is an extreme and unlikely high
CH$_4$ abundance---the hot atmospheric temperatures favor CO as
the dominant reservoir of C.

\clearpage

\begin{figure*}
\epsscale{1.0}
\plotone{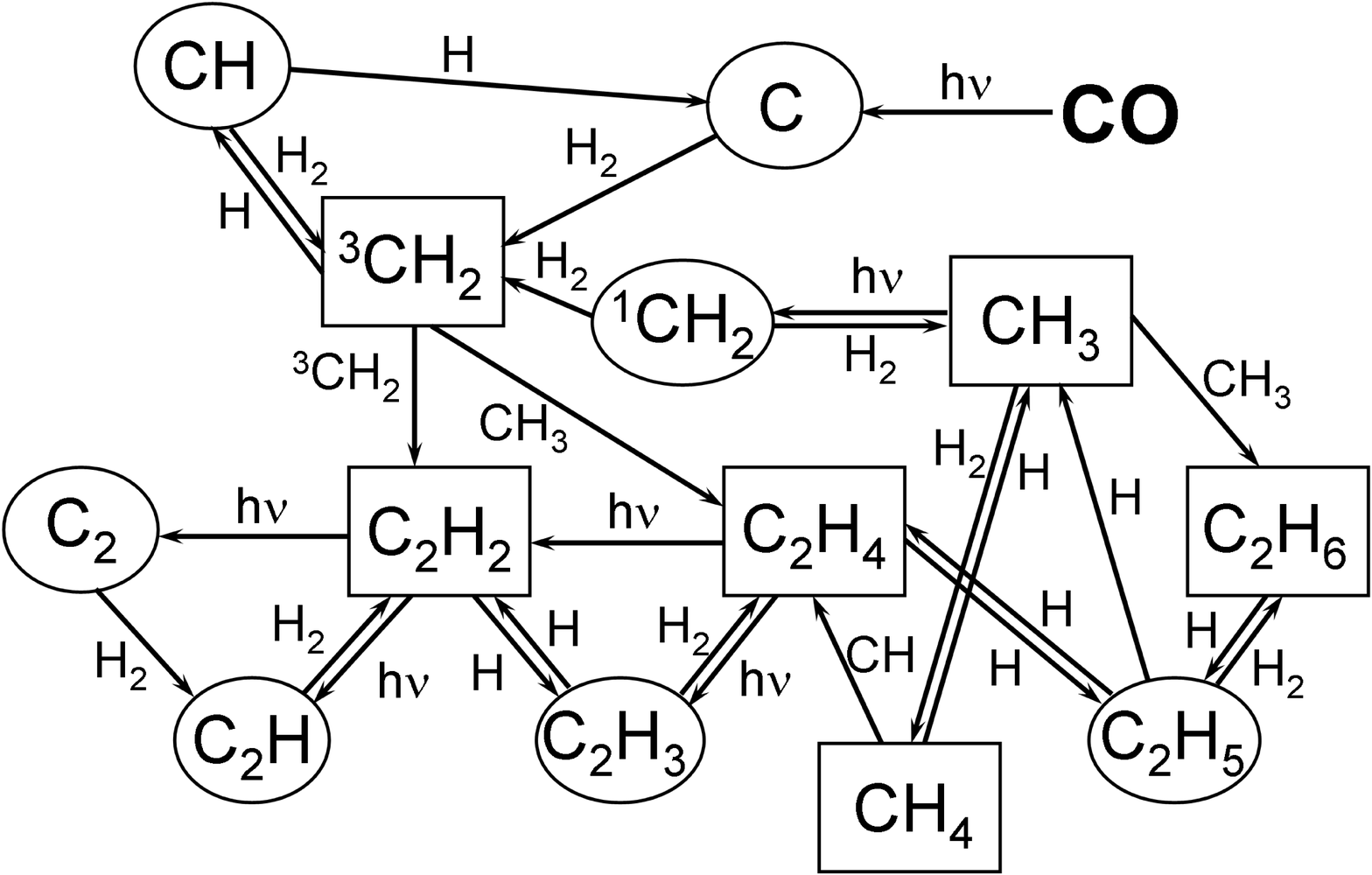} \caption[major pathways of C and
C2]{Major photochemical pathways for forming C and C$_2$ species.
\label{cc2}}
\end{figure*}

\clearpage

\begin{deluxetable}{rlllll}
\tabletypesize{\scriptsize} \tablecolumns{6} \tablecaption{Mixing
Ratios of CH$_4$, C$_2$H$_2$, C$_2$H$_4$, and C$_2$H$_6$ for
Models A-E at 0.1~mbar. Jupiter's Results at 2~$\mu$bar are
Included for Comparison. \label{column}} \tablewidth{0pt}
\tablehead{ \multicolumn{1}{c}{Model\tablenotemark{a}} &
\multicolumn{1}{c}{CH$_4$} & \multicolumn{1}{c}{C$_2$H$_2$} &
\multicolumn{1}{c}{C$_2$H$_4$} & \multicolumn{1}{c}{C$_2$H$_6$} &
\multicolumn{1}{c}{Total\tablenotemark{b}} }
\startdata
Jupiter                                       &  $1\times 10^{- 3}$  &  $1\times 10^{- 5}$  &  $3\times 10^{- 8}$  &  $2\times 10^{- 5}$  &  1      \\
A~1                                           &  $3\times 10^{- 6}$  &  $8\times 10^{- 7}$  &  $3\times 10^{- 8}$  &  $2\times 10^{-11}$  & $7\times 10^{-4}$ \\
\tablenotemark{d}~A~1                         &  $2\times 10^{- 6}$  &  $6\times 10^{- 7}$  &  $3\times 10^{- 8}$  &  $1\times 10^{-11}$  & $7\times 10^{-4}$ \\
\tablenotemark{e}~A~1                         &  $4\times 10^{- 6}$  &  $9\times 10^{- 7}$  &  $4\times 10^{- 8}$  &  $2\times 10^{-11}$  & $1\times 10^{-3}$ \\
\tablenotemark{f}~A~1                         &  $5\times 10^{- 7}$  &  $1\times 10^{- 7}$  &  $4\times 10^{- 9}$  &  $1\times 10^{-12}$  & $1\times 10^{-4}$ \\
B~1                                           &  $9\times 10^{- 6}$  &  $2\times 10^{- 6}$  &  $1\times 10^{- 7}$  &  $6\times 10^{-11}$  & $2\times 10^{-3}$ \\
C~1                                           &  $5\times 10^{- 7}$  &  $1\times 10^{- 7}$  &  $4\times 10^{- 9}$  &  $1\times 10^{-12}$  & $7\times 10^{-5}$ \\
D~1                                           &  $3\times 10^{- 5}$  &  $5\times 10^{- 6}$  &  $5\times 10^{- 7}$  &  $3\times 10^{-10}$  & $7\times 10^{-3}$ \\
E~1                                           &  $3\times 10^{- 4}$  &  $2\times 10^{- 5}$  &  $6\times 10^{- 6}$  &  $4\times 10^{- 8}$  & 0.4     \\
E~2                                           &  $2\times 10^{- 4}$  &  $9\times 10^{- 6}$  &  $1\times 10^{- 5}$  &  $1\times 10^{- 9}$  & 0.3     \\
E~3                                           &  $4\times 10^{- 4}$  &  $2\times 10^{- 5}$  &  $1\times 10^{- 5}$  &  $9\times 10^{- 9}$  & 0.6
\enddata
\tablecomments{The hydrocarbons have maximum mixing ratios at
about 0.1~mbar in the atmosphere of HD~209458b, while on Jupiter
the maxima are at about 2~$\mu$bar (see Figure~\ref{mr}).}

\tablenotetext{a}{1$:$ reference temperature profile \citep{SS00}.
2$:$ \citet{Betal02} temperature profile. 3$:$ \citet{Fetal03}
temperature profile.}

\tablenotetext{b}{Total$:$ column integrated abundances of
hydrocarbons (C$_2$H$_2$ + C$_2$H$_4$ + C$_2$H$_6$) at $< 2$~bar.
The abundance is normalized to $2\times 10^{-7}$ which is the
value calculated in the atmosphere of Jupiter (e.g., Gladstone et
al.  1996).}

\tablenotetext{d}{Exponent of eddy-diffusion is taken to be 0.65.}

\tablenotetext{e}{Eddy-diffusion is a factor of two smaller than
the reference eddy-diffusion.}

\tablenotetext{f}{Eddy-diffusion is a factor of ten greater than
the reference eddy-diffusion.}

\end{deluxetable}

\clearpage

\clearpage

\section{Discussion \label{discussion}}
Using a simplified version of the Caltech/JPL KINETICS model, we
have shown that the concentrations of the C$_2$H$_{2n}$ species
(see Table~\ref{column}) are insignificant in the atmospheres of
CEGPs. These C$_2$H$_{2n}$ compounds are important sources for
forming more complex C$_x$H$_y$ species, such as benzene and PAHs,
which will lead to the formation of hydrocarbon aerosols (e.g.,
Richter \& Howard 2000, 2002). Although we have used a simplified
photochemical model that captures the main reactions, we have
tested Models A-E using the reference temperature profile
(solid-line in Figure~\ref{profile}) incorporating the full
version of hydrocarbon model by \citet{Getal96}. Even for this
case, we find that the C$_6$H$_6$ abundance for Model A is seven
orders of magnitudes less than that of Jupiter and is two orders
of magnitudes less for Model~E. Sulfur and nitrogen containing
compounds are other potential sources for aerosols and we plan to
explore their photochemistry in a later paper.

The CEGPs are extremely close to the parent star; in such an
extreme environment, the C$_x$H$_y$ compounds will be lost either
primarily by reactions with atomic hydrogen or also by photolysis.
The production of atomic hydrogen is a consequence both of the
high temperatures that allow the presence of H$_2$O vapor and of
the high UV flux that causes photolysis of H$_2$O.  Therefore, the
lifetime of the C$_x$H$_y$ compounds in the atmospheres of CEGPs
is predicted to be much shorter than that on Jupiter.  The
lifetimes of the hydrocarbons are $\lesssim 10^3$~s, which are
significantly shorter than the simulated circulation timescale of
$\sim$day \citep{SG02,Cetal03}. Hence the abundances of the
hydrocarbons will be affected by a factor of `a few' through the
relatively longer lifetime of the atomic hydrogen ($\sim$1 day,
Liang et al. 2003).

The condensation temperatures for hydrocarbons (e.g., C$_4$H$_2$
and C$_4$H$_{10}$) are below 200~K at $\sim$1~mbar (Moses et al.
2000). These temperatures are far colder than expected in the
atmospheres of CEGPs \citep{SS00,Betal02,Fetal03}. Nevertheless,
we verified this by considering the saturation profiles together
with the the temperature profiles and found that the required
saturation pressure for CEGPs is far more than that present in the
atmospheres.

Using the measured Rayleigh scattering cross sections of He and
H$_2$ \citep{CD65,FB73}, the pressure level with optical depth
unity is $\sim$1~bar at 300~nm and increases rapidly at longer
wavelengths (Rayleigh scattering cross section $\propto
\lambda^{-4}$). Without the shielding from the atmospheric
aerosols and in the absence of high-temperature condensate clouds,
we may be able to observe the atmospheric composition at short
wavelengths up to the Rayleigh scattering limit.

In this letter, we have emphasized photolytically driven processes
involving neutral species. We have not considered the possibility of
ion-neutral chemistry, such as that found in the polar region of
Jupiter \citep{Wetal03}. This may be important in the atmospheres of
CEGPs if the planet possesses a magnetic field. If the hydrocarbon
aerosols can be formed in the polar region, then global circulation
will redistribute them to lower latitudes.  Stellar wind may be
another source of energetic charged particles that could result in the
formation of aerosols. Another subject not addressed in this work is
the formation of aerosols by heterogeneous nucleation in the presence
of pre-existing solid dust grains. In this case, the formation of
aerosols would be sensitive to the amount of dust particles in the
atmosphere.

Additionally, we find that the mixing ratios of C, O, S, and
C$_2$H$_2$ (other than H) are high at the top of the atmosphere,
implying that these particles can readily escape. The recent
detection of C and O in the extended upper atmosphere of
HD~209458b by \citet{Vetal04} supports this assertion and we
comment that hydrodynamically escaping atmospheric species will
yield new information on the evolution of CEGPs.

\acknowledgements We thank M. Gerstell, J. McConnell, and R. L.
Shia for helpful discussions. We thank an anonymous referee for
constructive comments. The support of NASA Grant NAG5-6263 to the
California Institute of Technology is gratefully acknowledged.
This material is also based upon work supported by the National
Aeronautics and Space Administration through the NASA Astrobiology
Institute under Cooperative Agreement No. CAN-00-OSS-01 and issued
through the Office of Space Science. S.S. is supported by the
Carnegie Institution of Washington and by NASA Origins grant
NAG5-13478.

\end{document}